\documentclass[%
 reprint,
superscriptaddress,
 amsmath,amssymb,
 aps,
pra,
]{revtex4-2}

\usepackage{graphicx}
\usepackage{dcolumn}
\usepackage{hyperref}
\usepackage{chemformula}
\usepackage[T1]{fontenc}

\begin{document} 


\title{Log-normal glide and the formation of misfit dislocation networks in heteroepitaxial ZnS on GaP}

\author{Alexandra Fonseca Montenegro}
\affiliation{Department of Materials Science and Engineering, The Ohio State University, Columbus OH 43210, USA} 

\author{Marzieh Baan} 
\affiliation{Department of Materials Science and Engineering, The Ohio State University, Columbus OH 43210, USA} 

\author{Maryam Ghazisaeidi}
\affiliation{Department of Materials Science and Engineering, The Ohio State University, Columbus OH 43210, USA} 
\affiliation{Department of Physics, The Ohio State University, Columbus OH 43210, USA}

\author{Tyler J. Grassman}
\affiliation{Department of Materials Science and Engineering, The Ohio State University, Columbus OH 43210, USA} 
\affiliation{Department of Electrical and Computer Engineering, The Ohio State University, Columbus OH 43210, USA} 
\affiliation{Center of Electron Microscopy and Analysis, Columbus OH 43210, USA} 

\author{Roberto C. Myers} 
\email{myers.1079@osu.edu} 
\affiliation{Department of Materials Science and Engineering, The Ohio State University, Columbus OH 43210, USA} 
\affiliation{Department of Physics, The Ohio State University, Columbus OH 43210, USA} 
\affiliation{Department of Electrical and Computer Engineering, The Ohio State University, Columbus OH 43210, USA} 


\begin{abstract} 
Scanning electron microscopy (SEM) based electron channeling contrast imaging (ECCI) is used to observe and quantify misfit dislocation (MD) networks formed at the heteroepitaxial interface between ZnS and GaP grown by molecular beam epitaxy (MBE). Below a critical thickness of 15-20 nm, no MDs are observed. However, crystallographic features with strong dipole contrast, consistent with unexpanded dislocation half-loops, are observed prior to the formation of visible interfacial MD segments and any notable strain relaxation. At higher film thicknesses (20 to 50 nm), interfacial MD lengths increase anisotropically in the two orthogonal in-plane <110> line directions, threading dislocation (TD) density increases, and a roughening transition is observed from atomically smooth two-dimensional (2D) to a multi-stepped three-dimensional (3D) morphology, providing evidence for step edge pinning via surface terminating dislocations. The ZnS strain relaxation, calculated from the total MD content observed via ECCI, matches the average strain relaxation measured by high-resolution x-ray diffraction (HRXRD). The MD lengths are found to follow a log-normal distribution, indicating that the combined MD nucleation and TD glide processes must have a normal distribution of activation energies. The estimated TD glide velocity ($v_{g}$) along [$\bar{1}$10] is almost twice that along [110], but in both directions shows a maximum as a function of film thickness, indicating an initial burst of plasticity followed by dislocation pinning. 
\end{abstract} 

\maketitle

\section{Introduction} 

At the coherent interface between an epitaxial thin film and substrate, the mismatch strain, $f = (a_{s}-a_{f})/a_{f}$, adds an interfacial energy growing linearly with film thickness ($h$), which can be relieved by dislocation-induced strain relaxation above some critical thickness, $h_{c}$. In closely lattice-matched systems, segments of dislocation loops residing at the film/substrate interface that contain some edge character, otherwise known as misfit dislocations (MD), provide a reduction of $|f|$ and its associated strain energy. These dislocations typically do not extend across the entire sample, but instead terminate via threading dislocation (TD) segments that intersect the top surface of the film and/or the bottom surface of the substrate. In the Matthews-Blakeslee model,\cite{matthews1974} shearing of a pre-existing TDs across the film-substrate interface results in MD segments once sufficient strain energy has built up, thereby defining $h \ge h_{c}$. However, in many cases the density of pre-existing TDs is insufficient to account for the observed MD content. These MDs are hypothesized to form via the nucleation of dislocation half-loops at the surface of the film, which propagate downward until the loop intersects the interface.\cite{hull1992} The generated MD segment grows in length as the two surface terminating TD segments glide parallel to the interface. The nucleation of MD segments, TD glide, and the resulting formation of a dislocation network in epitaxial films has been the subject of a large body of theoretical work\cite{jain1997,freund&suresh}. Many of the core predictions or assumptions of these theories have yet to be fully experimentally validated since dislocation imaging in epitaxial systems has historically been largely based on transmission electron microscopy (TEM), which require time-intensive and destructive specimen preparation, which may irreversibly alter subtle features, such as dislocation half-loops. Additionally, achievement of statistically relevant sample sizes via TEM, sufficient to adequately characterize the MD / TD networks to reveal the underlying formation pathways and dislocation dynamics, requires effectively heroic efforts. 

More recently, scanning electron microscopy (SEM) based electron channeling contrast imaging (ECCI), has emerged as a tool for rapidly imaging defects (dislocations and any other crystallographic defects) over large areas on as-grown, unprocessed specimens.\cite{carnevale2015,boyer2021,picard2014,picard2007,crimp2006} When combined with quantitative image analysis, statistically useful data sets can be acquired on defect populations in epitaxial layers to unveil the details of MD network evolution. Because TDs are generally found to be detrimental to (opto)electronic properties and performance, the typical goal of such efforts is to reduce the density of TDs by maximizing MD glide length, thereby minimizing the total number of dislocation loops required to achieve relaxation. However, it is also possible to consider dislocations as electrically active components themselves. Like two-dimensional electron gases (2DEGs) at the interface between narrow and large band gap semiconductors, MDs are localized at the heteroepitaxial interface where they formed during growth. The total length of all MDs ($\sum L_{MD} $) in a given area ($A$) defines their density, $\rho_{MD} [\mu m^{-1}] = \sum L_{MD}/A$, which grows with epilayer thickness that can be controlled with single atomic layer precision via growth techniques, like MBE. The generated MD networks can be characterized using ECCI at high-throughput and over large-areas, non-destructively mapping and quantifying the MD content on as-grown wafers before subsequent device processing. Demonstration of dislocations as electronically-active features was experimentally pioneered in Si,\cite{kittler2009} and more recently theoretically predicted in diamond\cite{genlik2023}. The narrow band gap of the former makes it challenging to electronically or optically identify signatures of dislocations, while the supreme mechanical strength of the former make it nearly impossible to glide dislocations via strain relaxation.   
 
We propose that the zinc-blende ZnS/GaP heteroepitaxial system provides a suitable test-bed for exploring the electronic properties of MDs since (i) the wide room temperature band gap of the ZnS epilayer (3.725 eV)\cite{Bagayoko2016} and GaP substrate (2.25 eV) increase the sensitivity to sub-bandgap levels associated with dislocations, (ii) ZnS is already known to exhibit electronic or optically active dislocations,\cite{ahlquist1972,oshima2018,Li2023} (iii) with $a_{ZnS} = 0.5420$ nm and $a_{GaP} = 0.5451$ nm, a small $f=0.57\%$ (MD extra-half plane within ZnS) ensures viable $h$ control of $\rho_{MD}$, and (iv) ZnS exhibits surprising plasticity at room temperature due to its extremely low stacking faulty energy ensuring that TD glide can occur during growth.\cite{takeuchi1985}

The wide and direct band gaps of ZnS and ZnS$_{1-x}$Se$_{x}$ made them popular candidates for blue LED development \cite{Summers1997}, until the advent and dominance of GaN/InGaN optoelectronics.\cite{nakamura_nobel_2015} As a result, heteroepitaxial development in ZnS was largely abandoned. Previous work investigated the epitaxial growth of ZnS with various substrates, including Si, sapphire, GaAs, GaN and GaP \cite{Schetzina1992,Horn1991,McGill1998,McGill1997,Kukimo1986}. ZnS grown on GaP was studied using Matthews and Blakeslee's force balance model estimating  $h_{c}=15$ nm \cite{Choi1998}. Experimentally, ZnS/GaP epitaxy was studied via HRXRD rocking curve linewidth, spectroscopic ellipsometry, resonant Raman scattering and high resolution transmission electron microscopy (HRTEM) along with reciprocal space maps (RSM) with resulting estimates for $h_{c}=21-35$ nm.\cite{Choi1998,Choi2002,Yamauchi2012} These characterization methods probe the onset of MD formation either indirectly, via the strain, or directly by imaging (TEM), but cannot readily quantify, with statistically relevant accuracy, the low densities of defects expected in the initial stages of MD generation.   

Here we study the early onset of strain-relaxation and MD network formation in heteroepitaxial ZnS grown on GaP (001) by molecular beam epitaxy (MBE). ECCI imaging is used to obtain large data sets to enable determination of MD length ($L_{MD}$) and density ($\rho_{MD}$) anisotropy for films grown at different $h$. We discuss the relationship between the formation and evolution of the dislocation networks with the strain state (HRXRD), surface morphology and growth mode (AFM and RHEED).  

 \section{Experimental Methods} 
 
ZnS epilayers are grown on a Veeco 930 MBE system with a base pressure of 10\textsuperscript{-9} Torr. The system is equipped with a Knudsen-cell ZnS compound source heated at the tip to 950 °C, which provides a flux measured as a beam equivalent pressure (BEP) of 1.50 × 10\textsuperscript{-6} Torr. Growth on as-received GaP (001) substrates using a variety of oxide removal methods (etching, in-situ oxide desorption) leads to roughening observed by RHEED (15 keV), and results in poly-crystalline or amorphous films. To address this issue, a two-chamber MBE method is employed. First, a buffer layer is grown homoepitaxially on as-received 2” semi-insulating GaP (001) substrates after oxide desorption at 800 °C and is followed by deposition of an arsenic capping layer in a separate III-As/P MBE system. The wafer is cleaved into quarters and then loaded into the ZnS MBE chamber. After the As-cap is desorbed, a smooth 2D surface is observed by RHEED. The substrates are heated to the growth temperature of 150 °C and ZnS is deposited at a rate of 1.05 $\mu m/hr$ with thicknesses of 15, 20, 25, and 50 nm. The surface morphology and structural composition of the layers is studied by atomic force microscopy (Bruker Icon 3 AFM ), high-resolution X-ray diffraction (Bruker D8 HRXRD), and scanning transmission electron microscopy (Titan 60-300 TEM) at an accelerating voltage of 300 kV and camera length of 460 mm. The HRXRD diffraction patterns are simulated and fit in LEPTOS software using two free parameters (thickness, h, and relaxation, R). Misfit dislocations and other defects present in the samples are observed with electron channeling contrast imaging (ECCI) conducted in an environmental scanning transmission microscope equipped with a back-scatter electron detector (Thermo Scientific Quattro ESEM) at an accelerating voltage of 30 kV, a current of 2.4 nA, and a dwell time of 8 $\mu s$. ECCI images were processed and quantified using MIPAR software. 

\section{Results and Discussion} 

The RHEED pattern along the [110] direction of GaP (001) substrate with a homoepitaxial buffer layer after arsenic desorption at 350 °C is shown in Figure 1(a). After 50 nm of ZnS growth the pattern transforms from spots to modulated streaks. The transition on the surface morphology from a flat surface to a multilevel stepped surface is indicative of a roughening transition within this 2D growth mode, which at thicker films transitions to 3D-islands in agreement with the literature\cite{Cavenett2001}. HRXRD taken along the (001) axis are plotted in Figure 1(b). 
\begin{figure*} 
  \includegraphics{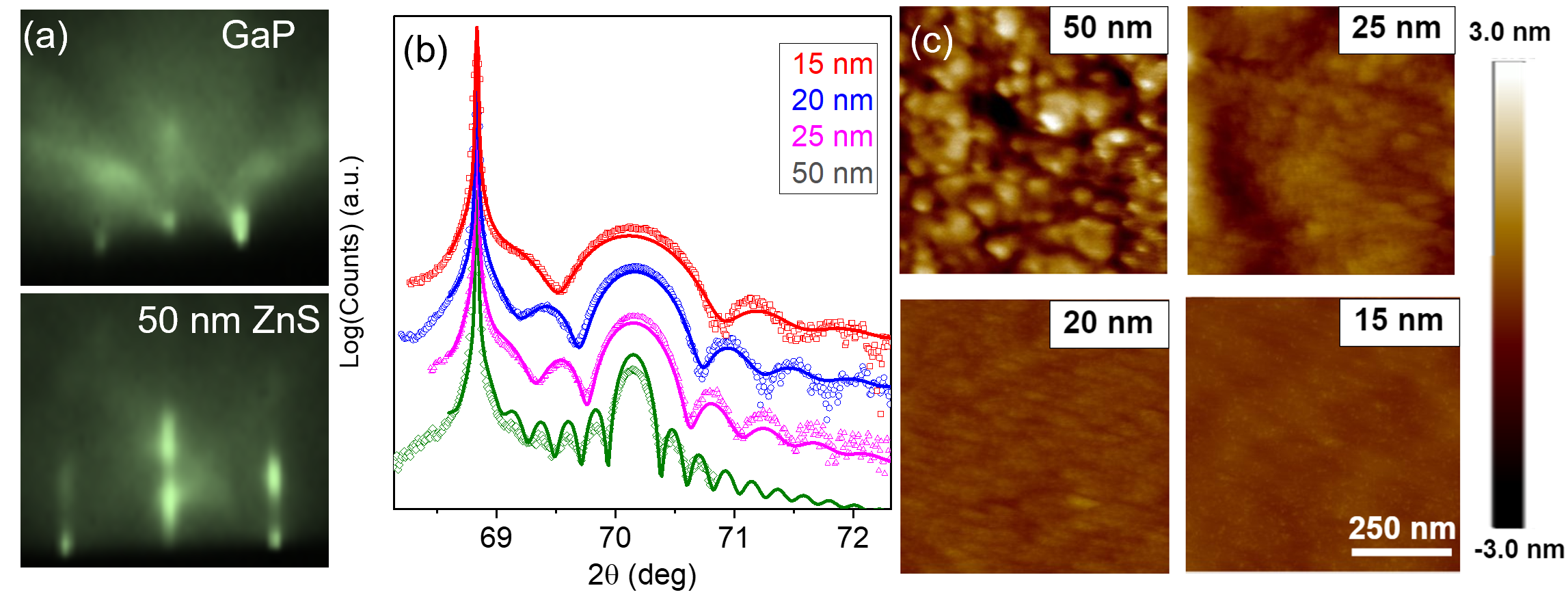} 
 \caption{Characterization of surface morphology and strain state of epitaxial ZnS grown on GaP (001) by molecular beam epitaxy (MBE).(a) In-situ reflection high energy electron diffraction (RHEED) along [110] of (top) the GaP surface just prior to ZnS deposition, and (bottom) just after deposition of 50 nm ZnS film. (b) High-resolution x-ray diffraction (HRXRD) data (points) and dynamical diffraction fit (lines) for determining film thickness ($h$) and relaxation ($R$). (c) Morphology of epilayers at various thicknesses by atomic force microscopy (AFM).} 
  \label{characterization} 
\end{figure*}
\begin{figure} 
\includegraphics{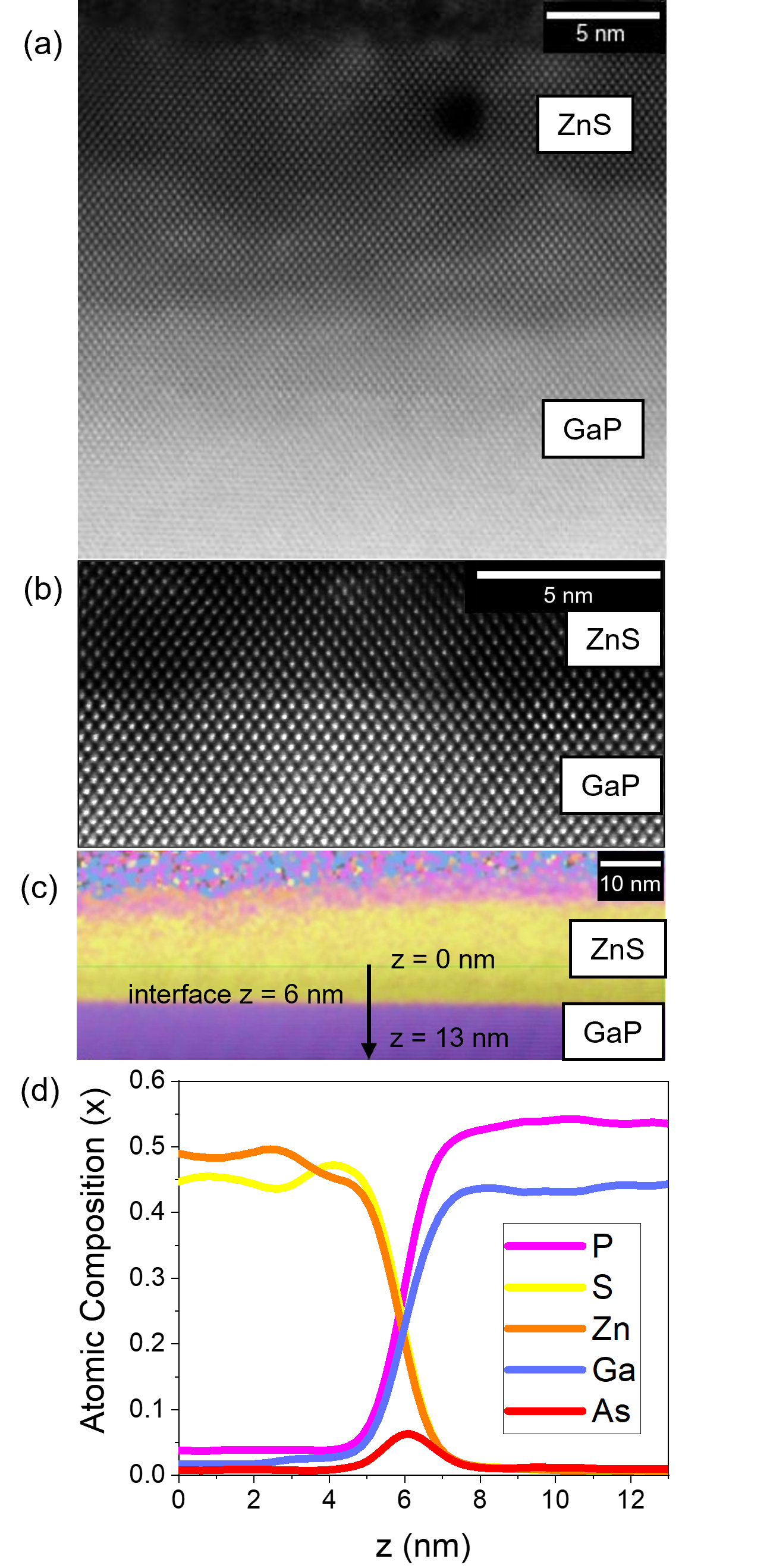} 
 \caption{(a)  Atomic resolution with Z-contrast (S/TEM HAADF) cross-sectional image of 15 nm ZnS film along [110] and (b) zoomed view of the same data set at the interface. (c) Chemically resolved image (EDS), and (d) line-cut extracted in the direction of the arrow of (c).} 
  \label{TEM} 
\end{figure}
Dynamical HRXRD simulation and fitting assuming isotropic (tetragonal films) in-plane strain is used to estimate the average R and h, which is sensitive to the finite-thickness fringes. We obtain h = 49.51, 24.64, 20.44 and 15.16 nm, for films that are nominally h = 50, 25, 20 and 15 nm, with R of 7.2\%, 2.4\%, 1.3\% and 0\%, respectively. The surface morphology of these films, measured by AFM, is shown in Figure 1(c). The 15 and 20 nm samples show an atomically smooth 2D surface layer with RMS roughness $\approx$1 monolayer (ML), 1 ML = 0.271 nm , consistent with layer-by-layer growth. The roughness increases markedly with h, in the range $\approx$4-6 ML for the thicker films. This correlates with the RHEED observation of modulated streaks (multilevel stepped-2D surface). The mophology of these rougher films shows the apparent formation of multi-stepped islands that reduce in area as thickness increases, which would be consistent with an increase of surface step pinning sites with thickness. 

Cross-sectional S/TEM (HAADF) of the h = 15 nm ZnS epilayer is shown in Figure 2(a) illustrating a coherent layer with no visible defects. The atomic resolution captures the strain of the material at the interface. The composition of the films is analyzed with energy dispersive X-ray spectroscopy (EDS) mapping as shown in Figure 2(b), revealing arsenic doping at the interface of the film. This is present because of incomplete As-desorption from the GaP substrate before ZnS growth, likely because of the finite solubility of As in GaP leading to an As-doped GaP surface. In Figure 2(c), the spectra for the EDS map shows that the atomic composition at the interface is 0.06 for arsenic.  

Defects are analyzed with ECCI. This technique follows the same principles as TEM in regards to the invisibility criteria, which is defined by $\vec{g} \cdot \vec{b}$ = 0 and $\vec{g} \cdot (\vec{b} \times \vec{u})$ = 0, where $\vec{g}$ is the diffraction condition, $\vec{b}$ is the Burgers vector, and $\vec{u}$ is the line direction \cite{Carter2009}. ECCI relies on low magnification electron channeling patterns (ECPs) to provide a map based on Bragg's condition that is used to orient the crystal. The sample is then tilted and rotated along the Kikuchi lines or bands to reach the desired diffraction vectors. Bright/dark contrast is generated from the localized strain fields by near-surface defects.  Figure 3(a) shows the ECP for ZnS/GaP films with circled region indicating the location on the ECP detector where ECCI micrographs are obtained by rastering the beam. 
 \begin{figure*} 
  \includegraphics{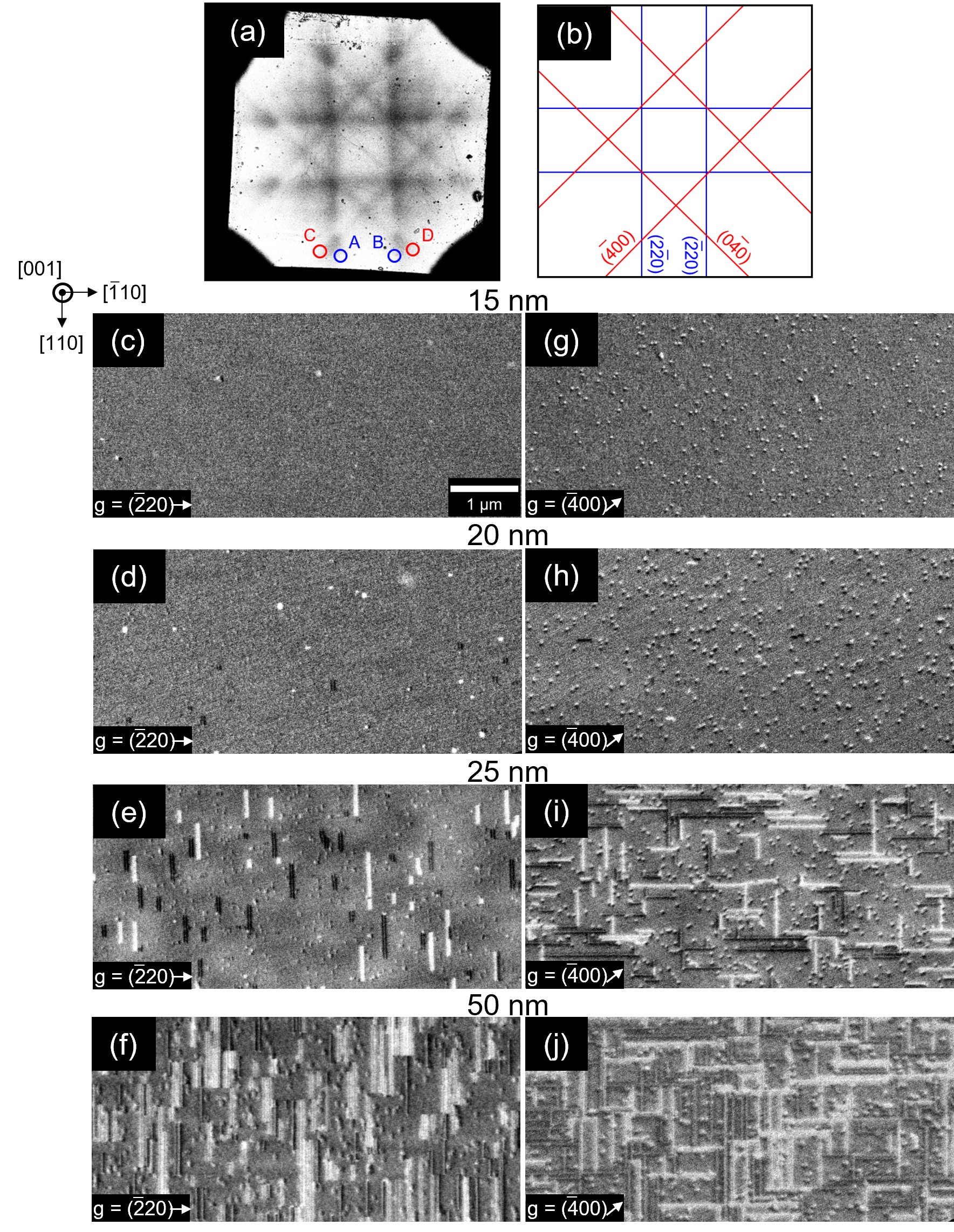} 
  \caption{Plan-view scanning electron microscopy (SEM) based imaging of misfit dislocation (MD) network formation in ZnS epilayers grown on GaP. (a) Electron channeling pattern (ECP) of ZnS/GaP and (b) indexed representation of Kikuchi lines. Electron channeling contrast imaging (ECCI) of ZnS/GaP films along diffraction vector, $g=$ ($\bar{2}$20) (c-f) and ($\bar{4}$00) (g-j) at each film thickness.}
  \label{ECCI} 
\end{figure*} 
 In Figure 3(b) the indexed channeling pattern shows A and B spots correspond to the (2$\bar{2}$0) and ($\bar{2}$20) Kikuchi lines, respectively. Similarly, the C and D spots contain the $\vec{g}=$ ($\bar{4}$00) and (0$\bar{4}$0) diffraction vectors, respectively. The electron micrographs taken from spot B ($\vec{g}=(\bar{2}20)$) are seen in Figure 3(c-f) showing either bright or dark contrast for line features. When imaged with  $\vec{g}$  of opposite sign at spot A of the ECP ($\vec{g}=(2\bar{2}0)$), the same line features reverse in contrast from bright to dark or vice versa (data not shown). This is the expected behavior if these are MDs with $\vec{u}=[110]$, but opposite sign of their out of plane Burgers vector components ($b_z$). For example, assuming the MDs to be $b<110>$ {111}, i.e. perfect 60$^{\circ}$, then for the given $\vec{u}=[110]$, the MDs may be from either (1$\bar{1}$1) or ($\bar{1}11$) slip planes. Since the direction of the extra half plane ($\vec{b} \times \vec{u}$)  must point to the top film ZnS film surface (not into the substrate) in order to relieve strain, then the possible Burgers vectors are either $\vec{b}=[0\bar{1}\bar{1}]$  or $[10\bar{1}]$ for the (1$\bar{1}$1) slip plane with half-plane along ($\vec{b} \times \vec{u}$) = [1$\bar{1}$1], or $\vec{b}=[01\bar{1}]$  or $[\bar{1}0\bar{1}]$ for the ($\bar{1}11$) slip plane with half-plane oriented along ($\vec{b} \times \vec{u}$) = [$\bar{1}$11]. Therefore, in the case shown in Fig. 3(c-f) diffraction of $\vec{g}=(2\bar{2}0)$ will be bright for the (1$\bar{1}$1) slip plane and dark for the ($\bar{1}11$), i.e. we can identify the out-of-plane component $b_z$ of the MDs. For the diffraction condition $\vec{g}=$ ($\bar{4}$00) shown in Fig. 3(g-j), MDs with both $\vec{u}=[110]$ or $[1\bar{1}0]$, provide diffraction contrast allowing precise quantification of the MD numbers and lengths in both directions. In addition to the MDs, small dipole features (DFs) can also be observed, i.e. bright-dark localized spots in the ECCI images. For thinner films such as the 15 and 20 nm samples, the DFs are not present for $\vec{g}=(\bar{2}20)$ in Figure 3(c) and Figure 3(d) until $h>h_{c}$. The DFs are consistent with the TD segments of surface nucleated dislocation half-loops imaged before they expand downwards into the material and glide at the interface to form a MD as illustrated in Figure 4(a).
 \begin{figure*} 
  \includegraphics{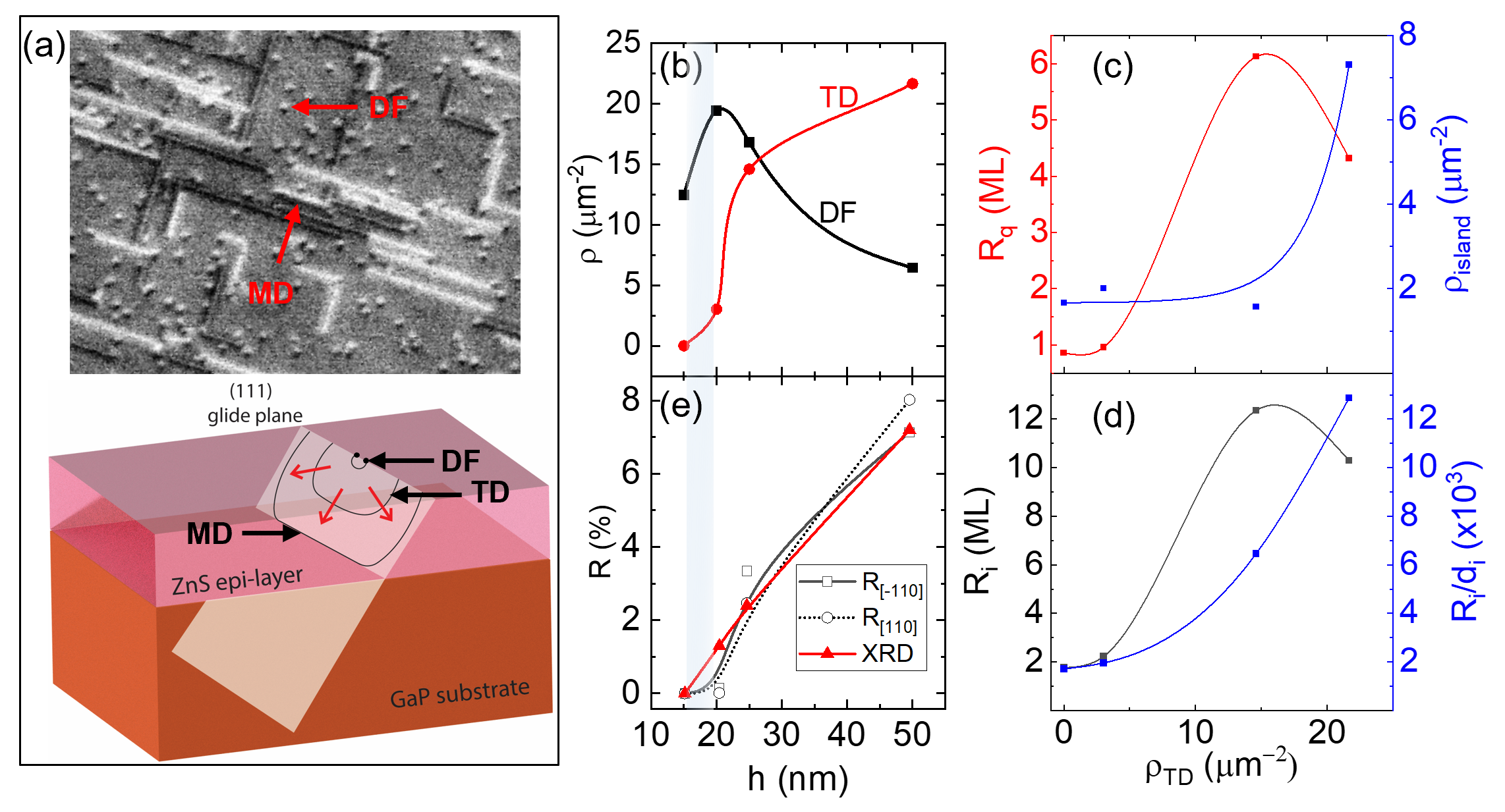} 
  \caption{Correlation between ECCI quantified defects, surface morphology and strain state. (a) ECCI image (Fig. 3(i)) of ZnS 25 nm thick epilayer on GaP identifying dipole features (DFs) and misfit dislocations (MDs) that are counted and measured in for every film thickness via image processing. The cartoon illustrates how the DFs are consistent with surface nucleated dislocation half loops from which MDs are thought to be generated. (b) Density ($\rho$) of DFs and threading dislocations (TDs) versus film thickness $h$. (c) AFM measured RMS roughness ($R_{q}$, red data) and island density ($\rho_i$, blue data) as a function of threading dislocation (TD) density ($\rho_{TD}$). (d) Average island height ($R_i$, black data) and average island aspect ratio ($R_i/d_i$, blue data) as a function of $\rho_{TD}$. (e) Strain relaxation state ($R$) of ZnS films as a function of $h$ determined from the ECCI measured MD density ($\rho_{MD}$) along  [110] and [$\bar{1}$10] line directions), and independently estimated from fits to the HRXRD data of  Fig. 1(b). Lines are guides to the eye}.
  \label{Relaxation}
\end{figure*} 
The DFs are oriented either along the [110] or [$\bar{1}$10] axes just as the MDs are, suggesting their related nature. Note that neither the MDs or DFs are observable on the secondary electron (regular SEM imaging) nor in the AFM confirming that the contrast is unrelated to surface morphology or contamination, and solely attributable to diffraction contrast proving their crystallographic nature, i.e. local strain-contrast. It is possible that they are stacking fault pyramids, which are observed in GaP/Si.\cite{feifel2020} Such features scale with film thickness, which does not appear to be the case, but could be obscured by the limited resolution and require S/TEM analysis.
 
No MDs are observed in the 15 nm sample in Figure 3(c) and Figure 3(g) confirming the fully strained layer independently measured via HRXRD (Fig. 1(b)), however for $h=20$ nm, small MD segments are observed (Figs.3(d) and 4(h)), therefore bracketing the critical thickness range, $15<h_{c}<20$ nm. As the thickness increases, MDs show a more diffuse contrast as seen on the 50 nm sample in Figure 3(f) and Figure 3(j). The strain field being captured by the ECCI gets scattered through the material as the film becomes thicker, making the bright and dark contrast from the dislocations more diffuse \cite{Grassman2014}. The defect data is extracted from the ECCI micrographs through MIPAR image processing software. The micrographs are adjusted in contrast and segmentation as described elsewhere \cite{Grassman2019}.  For each ZnS epilayer, the number of DFs, and the number and lengths of MDs are quantified based on 12 micrographs taken at 25k magnification at various spots along the samples and normalized by the imaged area. 

The DFs are observed even for $h<h_c$, showing a peak in density at 20 nm, near $h_{c}$, which decreases at larger $h$ (Fig. 4(b)). The apparent decay in DFs at increasing thickness might be a result of an undercounting of DFs due to their merging with the MDs. Yet, the trend in Fig. 4(b) is exactly what is expected if the DFs are indeed dislocation half loops, wherein the loops should increase in size with $h$, due to the increasing strain energy, but eventually decrease their density as the kinetic barrier for half-loop stabilization becomes larger as strain-relaxation proceeds via MD growth.\cite{hull1992} From the measured number of MDs, we also obtain the TD density ($\rho_{TD}$) if we assume that all the imaged MDs terminate at the surface through threading dislocations (TDs). While it is possible that some MDs form from pre-existing substrate TDs, the etch pit density (EPD) of the GaP substrates ($4.5\times 10^{-3}$ $\mu$m$^{-2}$) is negligible compared with $\rho_{TD}$  measured by ECCI ($\geq 2.5$ $\mu$m$^{-2}$). Comparing $\rho$ values of DFs with TDs in Fig. 4(b), we see further support for the identification of DFs as MD nuclei. TDs are not observed until  $h>h_c$, at which point the density of DFs peaks and is approximately $10\times$ that of the TDs, i.e. stable nuclei are present but only a small number have grown into MDs. At increased $h$ the TD density rapidly grows (plasticity burst) becoming comparable to the DFs, whose density has slightly decreased. The trend continues with additional TD nucleation and further DF reduction.

As previously discussed, the RHEED and AFM images (see Fig. 1) display a roughening transition indicative of an increase in surface step edge pinning with $h$. As the surface terminating TDs exert a local strain field, they alter the step-edge potential profile, and thereby can pin surface step edges, leading to roughening. The impact of surface terminating dislocations on adatom diffusion, and step advance has been studied both theoretically and experimentally in a number of materials.\cite{latyshev1995,heying1999,aminpou2011,oehler2013,schulz2022,miller2023} We empirically test this mechanism as a possible explanation for the roughening transition in ZnS/GaP by correlating the AFM morphology measurements with the ECCI based TD density measurements. Besides RMS roughness, ($R_{q}$),  image processing of the AFM images allows measurement of the average diameter ($d_i$), height ($R_i$), and density of islands  with height $>1$ ML ($\rho_i$), which are most pronounced in the 50 nm thick sample (Fig. 1(c)). $R_{q}$  and $\rho_i$ are plotted as a function of  $\rho_{TD}$ in Fig. 4(c) illustrating that from $R_{q}<1$ ML just at the onset of MD formation ($h\approx h_c$), with $\rho_{TD}\approx 3$ $\mu$m$^{-2}$, the surface transitions sharply to $R_{q}\approx 4-6$ ML as  $\rho_{TD}$ increases by $\approx 5\times$ concomitant with the appearance of islands of height $>2$ ML. The onset of roughening from an atomically smooth, unimpeded step-flow growth, to multi-step islands is positively correlated with a the burst of in $\rho_{TD}$. Assuming that the islands are formed by simply pinning the advance of their roughly circular boundaries via their interaction with the strain field of the TDs, one may expect that the stable island size will not become limited by TDs until a critical $\rho_{TD}$, since in the absence of pinning,  the stable island radius ($d_i/2$) is limited by the surface diffusion length of adatoms.  which is apparently in the range 140-260 nm. A critical $\rho_{TD}\approx 15$ $\mu$m$^{-2}$ is apparent in Fig. 4(c), where below this  density only a small number of multi-ML islands are observed, and above this critical density,  the island density approaches that of the TDs. If the TDs are limiting step-advance, leading to step-bunching, then we also expect an increase in the island height and reduction in their diameters. This is clearly observed in Fig. 4(d) where not only the height of the multi-ML islands increases, but the aspect ratio of the islands increases. Therefore, the independent AFM (film morphology) and ECCI  (crystallographic defects) measurements presented indicate TD-induced step edge pinning.

The ECCI quantified MD content appears to fully capture the strain-relaxation state of the films. This is illustrated in Fig. 4(d) where the ZnS epilayer relaxation ($R$) determined from dynamical diffraction fits to the measured by HRXRD (Fig. 1(b)) are plotted as a function of $h$. Independently, $R$ is calculated using the measured total MD length from ECCI imaging and determined $\rho_{MD}$. Assuming that all MDs are $\vec{b} = (a/2) <110>$, then the in-plane strain relief, is given as, $\delta_{[\bar{1} 1 0]}=(a_{ZnS}/4)*\rho^{[1 1 0]}_{MD}$ , where  $b_{eff}=|\vec{b}|/2 = a_{ZnS}/4$ is the edge component of the MD providing strain-relief of the film. Consequently the degree of relaxation along that direction is, $R_{{[\bar{1} 1 0]}}=\delta_{[\bar{1} 1 0]}/f\times 100\%$  from ECCI\cite{Grassman2021}. Figure 4(d) compares the HRXRD fitted strain relaxation with the ECCI based MD density estimate. There is quantitative agreement within the error expected for both HRXRD and ECCI. One notable discrepancy is the strain state in the 20 nm thick sample, where the ECCI based $R$ are $10\times$ smaller than the HRXRD estimate. The $\rho_{MD}$ would need to be nearly $100\times$ greater, very unlikely (see Fig. 3(d and h)) to match the HRXRD estimate. At these low $R$ values ECCI direct defect imaging is more accurate than HRXRD indirect two-parameter fit and strain-model. Overall, the ECCI measurements match the HRXRD strain estimates in terms of the thickness trend and absolute value of relaxation, i.e. within 1\%. This indicates that, unsurprisingly, the dominant form of strain-relaxation in the ZnS/GaP system is indeed from MD formation, and secondly, that ECCI is able to accurately quantify MDs. 

Besides the simple total MD density, $\rho_{MD}$, ECCI allows statistical analysis of the distribution of MDs and their anisotropy. The MD lengths are binned into histograms and plotted in Fig. 5(a).
\begin{figure} 

  \includegraphics{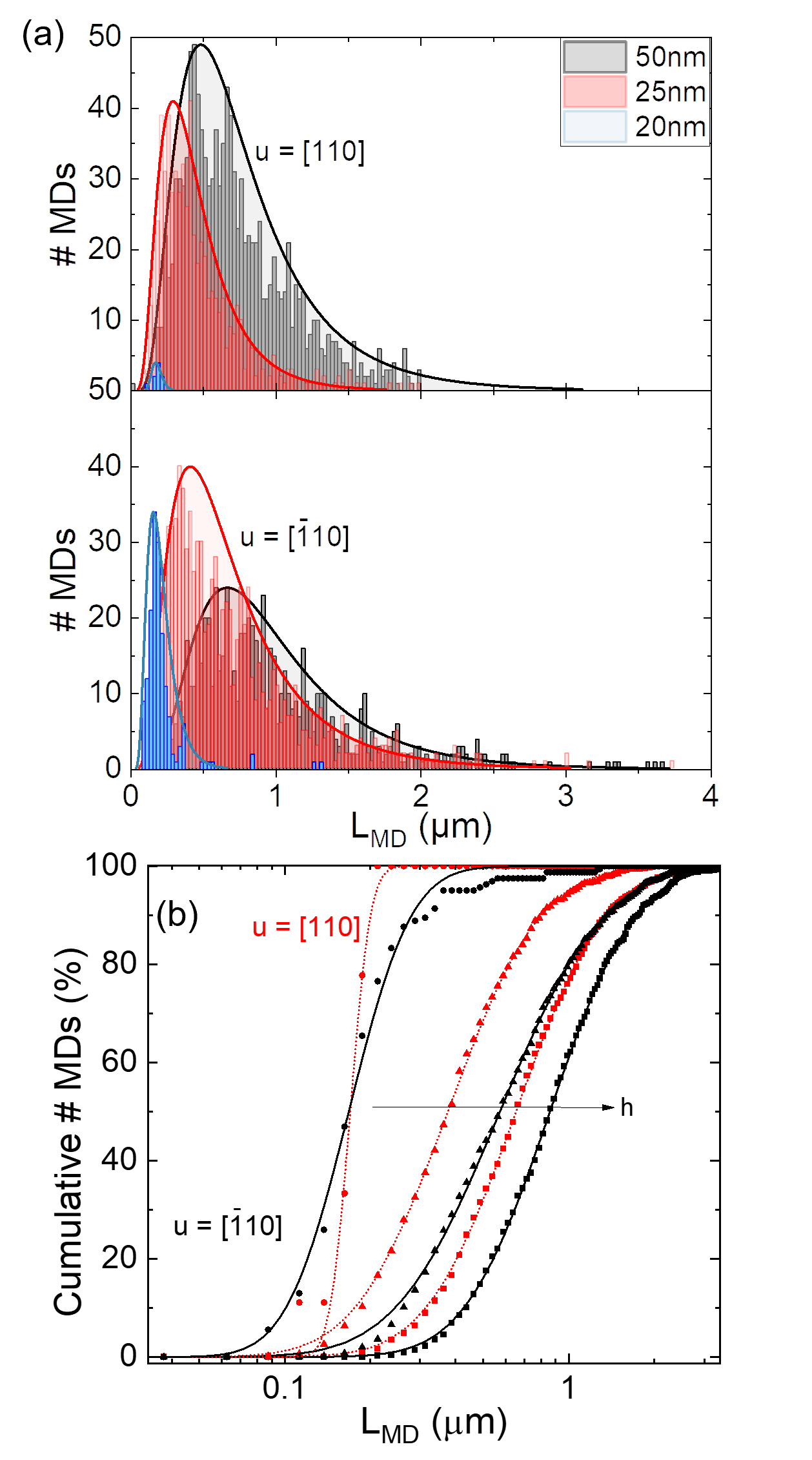} 

  \caption{Statistical analysis of misfit dislocation (MD) length measurements (see Fig. 3) for ZnS epilayers grown on GaP. (a) Number of MDs as a function of measured MD length ($L_{MD}$) binned every  25 nm. Data are plotted (bars) for three film thicknesses ($h$) and along both MD directions ($u$) with corresponding fits (lines) to a log-normal distribution. (b) The same data from (a) are replotted as a cumulative number of MDs plotted on a log-scaled x-axis, which enables two parameter fits of the data using the log-normal cumulative distribution function (CDF) plotted as lines.} 

  \label{Distribution} 

\end{figure}
The top and bottom panels show the variation in the distributions for the two line directions. The $L_{MD}$ anisotropy is most apparent near $h_c$ (at $h=20$ nm) favoring longer MDs along [$\bar{1}$10]. A shift in the distribution to longer $L_{MD}$ as TD glide proceeds is apparent. All six of the histograms fit well to the log-normal distribution, which emphasizes the importance of the highest percentile (tail of the distribution), i.e. the longest MDs. Note that fitting with other distributions (normal, Weibull, Poisson, etc.) all fail to provide agreement to the data set, while the log-normal captures clearly the entire data set. Similar behavior was reported in Si\textsubscript{1-x}Ge\textsubscript{x}/Si. \cite{Francesco2018} Fits to the raw histogram require three independent parameters, therefore to improve the accuracy of the fits we examine the cumulative $L_{MD}$ distribution, Fig. 5(b). These distributions are normalized (by the total MD length) such that all plots span 0 to 100\%, which reduces the number of fit parameters to two, $\mu$ and $\sigma$. The plot is shown with log-scaled horizontal axis, which converts the log-normal cumulative distribution function (CDF) to an apparently normal one. There is remarkably good agreement between the data (points) and (fits) to all six cases.

While the mean and standard deviation of the normal distribution have familiar intuitive meanings, it is difficult to similarly utilize the independent parameters $\mu$ and $\sigma$ of the log-normal distribution to characterize the MD population. Instead, we examine the $h$ dependence of the different percentiles of $L_{MD}$, plotted in Fig. 6, whose values are extracted from the CDF fits of the ECCI data, shown in Fig. 5(b).
\begin{figure} 

  \includegraphics{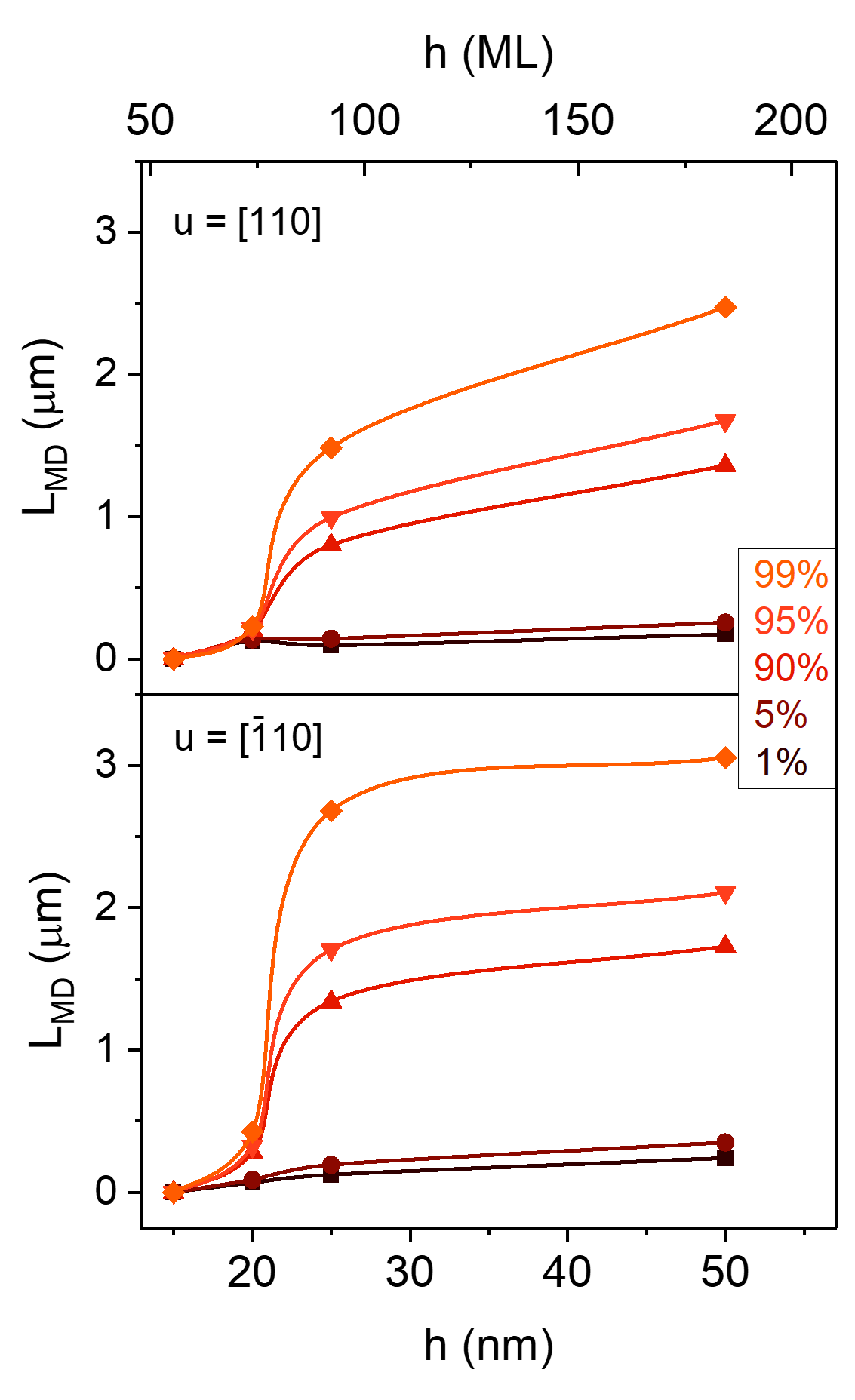} 
 
  \caption{Misfit dislocation (MD) length evolution with film thickness ($h$) based on log-normal cumulative distribution function (CDF) fits of the MD length ($L_{MD}$) data (shown in Fig. 3 and 5). The MD length at various representative percentiles is plotted, e.g. 99\% (1\%) represents the longest (shortest) population of MDs.} 

  \label{CF} 

\end{figure} 
 It is apparent that the  plasticity burst, previous discussed based on the TD density behavior shown in Fig. 4(b), is only apparent in the top >90\% long MDs. This could be entirely due to the majority of MD segments with quite limited length due to pinning of their surface terminating TD segments,\cite{Kunal2018} while the majority of the strain-relief is generated by a small population of long MDs generated by thermally activated TD glide. However, that interpretation assumes that all MDs nucleated at the same time, i.e. at $h=h_c$, ignoring the fact that the nucleation process itself is a thermally activated process. Thus, the $L_{MD}$ of any given MD is the product of the glide velocity ($v_g$) and the glide time ($t)$ minus the nucleation delay ($\Delta t_n$) before that MD first formed. A log-normal distribution of $L_{MD}$ is produced if the activation energy for glide ($E_g$) and/or nucleation ($E_n$) are normally distributed since $L_{MD}= v_g(t-\Delta t_n) \propto e^{(E_g/kT)}(t-e^{(E_n/kT)})$. This offers a plausible explanation for the observation of a specifically log-normal distribution in $L_{MD}$ for all film thicknesses and both <110> directions shown in Fig. 4. 

To interpret the $L_{MD}$ distribution and its evolution with thickness, we assume that the longest 99\% MDs in any sample have $\Delta t_n=0$, i.e. they were nucleated at $h=h_c$. The TD glide time ($t$) is assumed to be equal to the time between the first MD nucleation at $h_c$ and the end of the ZnS deposition. Therefore, only the longest 99\% of MDs are representative of the glide velocity ( $v_g$), which can be estimated under the assumption that all observed MDs are generated via surface nucleation of half-loops generating a single 60$^{\circ}$ MD segment and two TD segments traveling in opposite directions. This restriction in the kinetics of strain-relaxation is well supported, as discussed above, by the low EPD density of the starting substrates, and the agreement between the HRXRD and the ECCI estimated strain state of the ZnS epilayers. Figure 7 plots the resulting estimates for $v_g$ assuming various values of $h_c$ spanning the uncertainty as a function of film thickness normalized by the critical thickness ($h/h_c$). 
\begin{figure} 

  \includegraphics{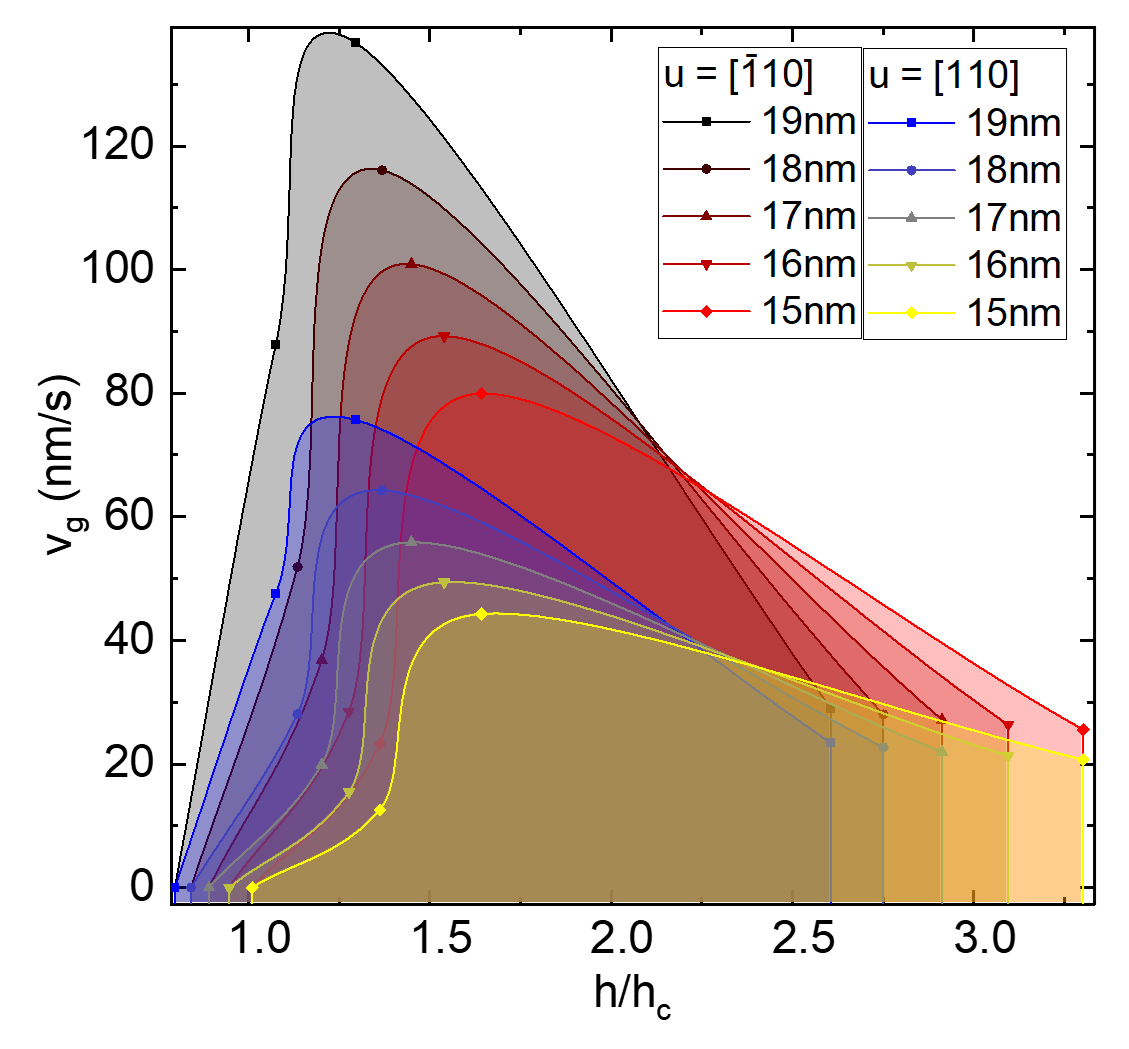} 

  \caption{Dislocation glide velocity ($v_{g}$) evolution with normalized film thickness ($h/h_{c}$) and in-plane glide direction ($u$) for ZnS epilayers grown on GaP. $v_{g}$ (points) are calculated from the MD length measurements and statistical analysis (see Figs. 3-6) assuming the longest (99\%) of MDs are nucleated exactly at  $h = h_{c}$. Data are plotted with $h_{c}$ varied over the its range of possible values (15-19 nm) to estimate the uncertainty in $v_{g}$. Lines are guides to the eye.}

  \label{Glide} 

\end{figure}
There is a roughly 2$\times$ greater $v_g$ for TDs along $[\bar{1}10]$ than along $[110]$, spanning the uncertainty in $h_c$. Anisotropy on the nucleation and glide velocity of MDs is well studied in III-V zincblende materials, which is attributed to the distinct core structure of the 60$^{\circ}$ <110> dislocations, $\alpha$ (III-core) and $\beta$ (V-core), with MD segments along $[\bar{1}10]$ and $[110]$, respectively.\cite{yonenaga1993,goldman1995,matragrano1996} 
Our ECCI-based $v_g$ data show that MD segments along $[\bar{1}10]$ are faster than those along the $[110]$, pointing to an anisotropy in the mobilities of the different core types, but more analysis is required to identify the core type along each line direction in our ZnS film.
Similar arguments were made earlier  by Brown, Russel, and Woods to explain the anisotropy of microtwin, cracks, and MDs observed in ZnSe/ZnS epilayers.\cite{brown1989} Additionally, for all values of $h_c$ two regions of glide are observed: (i) $1.0 < h/h_{c} < 1.5$, $v_{g}$ increasing with $h/h_{c}$, and  (ii) $1.5 < h/h_{c}$, $v_{g}$ decreasing with $h/h_{c}$. This is interpreted as (i) representing the region where the driving force for strain-relaxation growing linearly with $h/h_c$ (unimpeded TD glide), whereas region (ii) represents a suppression of TD glide despite the fact that $R<10\%$, such that the driving force for TD glide is only slightly reduced from that of region (i) implying that TDs are being pinned. The so-called "jerky motion" of dislocations was recently captured via ECCI in GaAs/Si epilayers by Callahan et al. revealing the repeated pinning of individual dislocations thought to be associated with TD-TD interaction\cite{Kunal2018}.

\section{Conclusions} 

SEM based ECCI quantification of defect content in epitaxial ZnS grown on GaP (001) enables a rich examination of the MD network nucleation and evolution kinetics. Before the the onset of MD generation, we observe the formation of DFs consistent with surface nucleation of dislocation half-loops, in support of long-standing theory. The evolution of film morphology with film thickness ($h$) measured by AFM and RHEED are correlated with ECCI quantification of dislocation content, showing a surface roughening and multi-step island formation associated with the burst in plasticity, as indicated by a rapid increase in TD density. The ECCI quantified MD content fully account for the average strain state of the films estimated from HRXRD measurements. Direct imaging of large data sets of MDs in both in-plane line directions provides suitable sample sizes for statistical analysis of the evolution of MDs as a function of $h$. Spanning more than two orders of magnitude, the distribution of MD lengths follows a strict log-normal distribution indicative of a normal distribution in the activation energy for nucleation and/or glide. Assuming that MDs evolves via TD glide, thickness dependence of $v_g$ is estimated accounting for the uncertainty in $h_c$. As the driving force for TD glide increases linearly with $h/h_c$, one would expect an increasing or constant $v_g$ until the strain is substantially relaxed. However, we find a rapid reduction in $v_g$ at  $h/h_c>1.5$ for $R<10\%$ indicative of TD pinning as recently observed in GaAs/Si.\cite{Kunal2018} These measurements establish a baseline for exploring patterning of MD network formation in ZnS using additional inputs, such as local strain  patterning,\cite{grydlik2013,mondiali2014}, optical excitation,\cite{ahlquist1972,oshima2018} or electronic biasing.\cite{Li2023}

\section{Acknowledgements} 

Financial support from the Air Force Office of Scientific Research (Grant FA9550-21-1-0278, Program Manager Dr. Ali Sayir) is acknowledged.

\bibliography{references} 

\end{document}